# New structural variation in evolutionary searches of RNA neutral networks.

**Sumedha** [1][§], **Olivier C. Martin**[1,2] **and Andreas Wagner**[3]

[1]  LPTMS, Université Paris-Sud, bât. 100, 91405 Orsay Cedex

[2]  INRA, UMR. de Génétique Végétale, Ferme du Moulon, 91190, Gif-sur-Yvette

[3]  Department of Biochemistry, University of Zurich
   Winterthurerstrasse 190, CH-8057 Zurich, Switzerland

[§] Corresponding author.

**Abstract**

RNA secondary structure is an important computational model to understand how genetic variation maps into phenotypic (structural) variation. Evolutionary innovation in RNA structures is facilitated by neutral networks, large connected sets of RNA sequences that fold into the same structure. Our work extends and deepens previous studies on neutral networks. First, we show that even the 1-mutant neighborhood of a given sequence (genotype) $G_0$ with structure (phenotype) $P$ contains many structural variants that are not close to $P$. This holds for biological and generic RNA sequences alike. Second, we analyze the relation between new structures in the 1-neighborhoods of genotypes $G_k$ that are only a moderate Hamming distance $k$ away from $G_0$, and the structure of $G_0$ itself, both for biological and for generic RNA structures. Third, we analyze the relation between mutational robustness of a sequence and the distances of structural variants near this sequence. Our findings underscore the role of neutral networks in evolutionary innovation, and the role that high robustness can play in diminishing the potential for such innovation.

**Introduction**

A properly formed RNA secondary structure is necessary for the biological functions of many RNA molecules, and a variety of algorithms exist to determine RNA secondary structure from an RNA sequence (Hofacker et al. 1994; Tacker et al. 1996; Zuker 2000). For these reasons, RNA secondary structure is an important computational model to understand how genetic variation maps into phenotypic (structural) variation (Fontana 2002; Fontana & Schuster 1998a; Schuster et al. 1994), and thus to understand the evolutionary dynamics of molecular innovations.

Our point of departure are RNA sequences that adopt a specific minimum free energy (mfe) secondary structure, which we can think of as necessary for some hypothetical biochemical process. This process might involve catalysis or just specific binding to some molecule. Some variant of this structure – perhaps very rare in the space of all possible structures – may greatly improve this biological function, or it may even lead to a new function. The question is how to find such a variant, *if we are not allowed to destroy the original structure* during an evolutionary search for this innovation. Part of the answer lies in the fact that the sequences folding into a given structure form one or a few "neutral networks" that can be traversed by single point mutations, and that span most of sequence space. This holds at least for generic structures, structures into which a sufficient number of sequences fold (Schuster et al. 1994).

An evolutionary search that must not destroy the original structure is effectively restricted to the 1-mutant neighborhood of a neutral network (Fontana & Schuster 1998b). (We define a *k*-mutant neighborhood of a sequence $G_0$ as containing all sequences that differ from $G_0$ in at most *k* residues.) Any such evolutionary search would start at one sequence. The 1-mutant neighborhood of this sequence contains only a limited number of structural variants, and thus only limited potential for evolutionary innovation. However, since the sequences folding into a structure are *connected* in a neutral network, the search can explore a great many sequences and their 1-mutant neighbors, without ever leaving the original structure (Schuster et al. 1994).

Among the substantial body of work that has explored the relation between sequence and structure space (Fontana & Schuster 1998b; Huynen 1996; Huynen et al.

1996; Reidys et al. 1997; Schuster et al. 1994; van Nimwegen et al. 1999), one paper (Huynen 1996) is of particular relevance. That paper focused on a specific, biologically important RNA structure, that of phenylalanine tRNA (tRNA$^{Phe}$). It showed that an exploration of this structure's neutral network through a random walk encounters an ever-growing repertoire of new structures in its neighborhood, a repertoire that does not become exhausted even for very long random walks. In addition, the 1-neighborhoods of distant sequences on the neutral network share very few structural variants. Another important result from previous work is that a thermodynamically stable and mutationally robust sequence encounters few structural innovations in its neighborhood (Ancel & Fontana 2000). We here extend and deepen these previous analyses. First, we show that continual structural innovation is a property not only of biologically important, but also of generic structures. Second, we statistically explore the relationship between mutational robustness and structural innovation, not only by counting the *number* of structural variants, but also by analyzing their *distances* to a reference structure. Importantly, we do so for generic structures, and not just for biologically important structures.

**Results**

*Around any one sequence, substantial structural variation is abundant.*

Consider a reference sequence (genotype) $G_0$ and its structure (phenotype) $P$. Among the 1-mutant neighbors of $G_0$, (there are $3n$ of them), some fraction will adopt a structure different from $P$. One would think that most structural variants will only differ slightly from $P$, because base-pair stacks resist structural changes in response to single base changes, be it for both biological and random RNA structures (Higgs 1993).

To find out whether this is the case, or whether a local exploration around a given sequence generates a great diversity of new structures, we took the following approach. We sampled randomly chosen genotypes $G_0$ with a given structure $P$, and determined the distribution of the *structure distance D* between $P$ and the structures of the 1-mutant neighbors of $G_0$ (see Methods for details). Specifically, we used two biological structures in this approach. The first structure is a 54-mer hammerhead motif of an RNA in peach latent mosaic viroid (Ambros et al. 1998). To ascertain that our results were not artifacts of the specific structure we chose, we also determined the same distance distribution averaged over many randomly chosen structures of length 54 (see Methods). The second biological structure was a phenylalanine tRNA (tRNA$^{Phe}$) of length 76, where we similarly determined for comparison the distance distribution for many randomly chosen 76-mers.

Figures 1a and 1b show these distance distributions. A key qualitative feature is similar for the two biological structures ($n$=54 and 76) and the random structures, in that the most probable distance is the smallest possible distance ($D$=2). Thus, variants of any one structure tend to be similar to it. However, we also note that *most* of the distributions' mass is located at moderate to large distances. For example, the median structure distances for the 1-mutant variants of the biological structures are $D$=14 (hammerhead) and $D$=24 (tRNA). This means that a typical single point mutation affects 7-12 base pairs, even in these short structures. All this is not just an artifact of a particular measure of structure distance: we also see it for the "bond" distance (see Methods; Figure 2), where median structure distances are $D$=12 (hammerhead) and $D$=22 (tRNA).

In sum, even at the smallest possible sequence distance, one observes a broad spectrum of structures with varying distance from a reference structure. Needless to say, however, the number of structures accessible from *anywhere* on a neutral network is much larger than that found in a single sequence's 1-neighborhood. For instance, for our 54-mer hammerhead motif we found 259689 distinct structures among $10^6$ randomly generated innovative 1-mutants of sequences on the neutral network. Furthermore, this number rises linearly with our sampling size, suggesting that the *total* number of distinct structures is many times larger still.

*New structural variation accumulates rapidly along a random walk of a neutral network.*

We next examine how rapidly structural variation accumulates during a random walk on a neutral network. Consider a reference sequence (genotype) $G_0$, its structure (phenotype) $P$, and the set $I_0$ of all structures that occur in the 1-neighborhood of $G_0$ and that are different from P. Denote by $G_k$ a sequence that differs at exactly k nucleotides from $G_0$, and that adopts the same structure P as $G_0$; denote by $I_k$ the set of structures found in the 1-neighborhood of $G_k$ that are different from P. The set $N_k = I_k \setminus (I_0 \cap I_k)$ is the set of *new* structures around $G_k$, that is, the set of structures that occurs in the 1-neighborhood of $G_k$, but not in the 1-neighborhood of $G_0$ (Figure 3). Then $X_k = |N_k|/|I_k|$ is the fraction of new structures around $G_k$, where |A| denotes the number of elements in a set A. As Figure 4 shows, new structures accumulate rapidly with increasing k. The greater the distance of a sequence $G_k$ on the neutral network from the starting sequence, the larger the fraction of new structural variation around it. This accumulation reaches a plateau at Hamming distances that are small compared to the diameter of sequence space (Figure 4), and it remains similarly flat for larger Hamming distance (data not shown). At this plateau, more than 80% of structures encountered at each step are new. This pattern is independent of whether random sampling or inverse folding (Hofacker et al. 1994) is used to generate sequences that fold into a given structure (Figures 4 and 5). (See methods for details on these two sampling methods.) We note parenthetically that if we do not restrict the evolutionary search to the neighborhood of a neutral network, but to any sequences at distance k from the reference sequence, then the fraction of new structural variants also reaches a plateau at similarly moderate values of k. However, at this plateau 100% of sequences encountered at any given k are new.

*The structure distance of new structures depends on k*

The previous analysis regarded the fraction of new structures in a 1-neighborhood of $G_k$, ignoring how different these structures are to each other and to the reference structure P. To ask how *structure distances* of new structures depend on k, we determined the average pairwise (Hamming) distance $D_{I_0, I_k}$ of all structures in the sets $I_0$ and $I_k$ (as defined in Figure 3). We also determined the average distance $D_{P,N_k}$ of structures in $N_k$ to the structure P itself. The results are shown in Figure 6 as a function of k. Qualitatively, the average pairwise (Hamming) distance $D_{I_0, I_k}$ increases with k. This means that the structures found in the 1-neighborhoods of $G_0$ and $G_k$ are increasingly different from each other as *k* increases. One might think that this would also hold for the distances of the *new* structures around $G_k$ to P. However, this is not so: $D_{P,Nk}$ falls precipitously with *k*. This means that starting with small *k,* the set of new structures around $G_k$ – structures in $N_k$ – is dominated by structures that are very dissimilar to P, but as *k* increases, their distance to P is increasingly dominated by structures similar to P.

This last observation may seem counterintuitive, but is easily explained. First note that if we take *all* structures around $G_k$ that are different from P (i.e, structures in $I_k$, not just structures in $N_k$.), then there is no dependence of $D_{P,Ik}$ on *k* at all when averaging over $G_0$. The reason is that $D_{P,Ik}$ is just the average distance to P of sequences in the 1-neighborhood of randomly chosen sequences on a neutral network. Now to understand the decrease of $D_{P,Nk}$ with *k*, we note that the structures in the 1-neighborhood of any sequence such as $G_0$ have a skewed distribution, with a small number of frequent structures and a larger number of rare structures. The frequent structures tend to be more similar to P than the rare structures. For small *k*, the set $N_k$ will comprise mostly rare structures, whose distance to P is, on average, greater than the distance of the frequent structures to P. As *k* increases, the sets $I_0$ and $I_k$ share fewer and fewer members, meaning that the set $N_k$ approaches $I_k$ itself in size. Thus, $N_k$ becomes increasingly dominated by frequent structures. Because these structures are closer to P than rarer structures in $N_k$, the pattern we observe follows. Qualitatively, the observations from Figure 6 also hold for the bond distance (Figure 7; see Methods), and are thus not sensitive to the specific distance measure used.

*The number and distance of new structures depends on mutational robustness.*

By definition, sequences $G_0$ that are highly robust to mutations, i.e., sequences that have many neutral neighbors, have *few* neighbors with innovative structures. This raises the possibility that these few innovative structures are structurally close to P, which is indeed the case. Figure 8 shows the relationship between mutational robustness (horizontal axes) and the average distance to P of its innovative 1-neighbors (vertical axes). The higher the robustness of $G_0$ the more similar are its non-neutral neighbors to P. Mutational robustness and thermodynamical stability are highly correlated, as has been previously noted (Ancel & Fontana 2000), and as we also observe for the sequences we study (results not shown). This means that thermodynamically stable sequences also have structurally more closely related variants.

We also asked whether the distances $D_{I_0, I_k}$ of structures in the sets $I_0$ and $I_k$ (as defined in Figure 3), and the distances $D_{P,N_k}$ of structures in $N_k$ to $P$ itself depend on the mutational robustness of $G_0$. To this end, we grouped our references genotypes $G_0$ into three groups of low, intermediate, and high robustness. We then plotted for each of these groups the distance measures as a function of $k$. Figure 9 shows the results for tRNA$^{\text{Phe}}$. Qualitatively, the dependency on $k$ is the same as that in Figure 6: $D_{I, I_k}$ increases with $k$, whereas $D_{P,N_k}$ decreases. The main difference is that for small values of $k$, $D_{I_0, I_k}$ and $D_{P,N_k}$ are lower when the sequence $G_0$ is highly robust.

**Discussion**

In sum, we here showed, first, that even the 1-neighborhood of a given sequence (genotype) $G_0$ with structure (phenotype) $P$ contains many structural variants that are not close to $P$. This holds for biological and generic (random) sequences alike. Second, as one walks further and further away from $G_0$ on its neutral network, the fraction of new structures, structures that are not already in the 1-neighborhood of $G_0$, increases rapidly, until it plateaus at a value of greater than 80%. This means that in the 1-neighborhoods of genotypes $G_k$ that are only a moderate distance $k$ away from $G_0$, a large proportion of structures have not been encountered before. This had been shown qualitatively for tRNA$^{\text{Phe}}$ (Huynen 1996). We here demonstrate it also for another biological structure and, importantly, show that it is not a peculiarity of biological structures, but a generic feature of RNA structures. Third, the structures in the 1-neighborhoods of $G_0$ and $G_k$ become increasingly distant from each other with increasing $k$. Fourth, and also relatedly to previous work (Ancel & Fontana 2000), we show that there are not only fewer new structures in the 1-neighborhood of a mutationally robust sequence $G_0$, but that these structures are also more similar to the structure $G_0$. This confirms that mutational robustness, although in moderation important for structural innovation, hinders such innovation at high values. We show that this is a generic feature of RNA structures, not only of a small subset of important RNA structures.

We note that in all our analyses we avoid coarse-graining, an approach that classifies secondary structures according to their number and sequence of stems and loops. Coarse-graining greatly reduces the complexity of structure space and thus makes its characterization more practical (Fontana & Schuster 1998b). However, two sequences with the same coarse-grained structure may have substantially different biological activities. For example, shortening of a stem in a catalytically active Hammerhead ribozyme leaves the coarse-grained structure intact, but it can lead to loss of catalytic activity (Birikh et al. 1997).

RNA and its secondary structures have two potentially important features for evolutionary innovation. First, as mentioned earlier, the sequences folding into a given structure form one or a few neutral networks that can be traversed by single point mutations, and that span most of sequence space. (Schuster et al. 1994). Second, even a small neighborhood around any one reference sequence can contain representatives of nearly all structures. Specifically, for sequences of length $n$, a neighborhood of radius $0.2n$ can contain representatives of nearly all structures. (The radius of a neighborhood is the maximal Hamming distance of any of its sequences from a reference sequence). In other words, representatives of nearly all structures can be found in a "ball" whose radius is only 20 percent of the radius $n$ of the entire sequence space. This property has been called *shape space covering* (Schuster et al. 1994).

The importance of neutral networks for evolutionary innovation is made clear by earlier analyses and by our work. But just how important is shape space covering for evolutionary innovation? Consider, for example, sequences of length $n=100$, whose sequence space comprises $4^n = 1.61 \times 10^{60}$ sequences because each base can be A,C,G or U. A ball of radius $0.2n=20$ comprises approximately $(3n)^{20}/20! = 1.43 \times 10^{31}$ sequences, which is only a fraction $8.9 \times 10^{-30}$ of the entire sequence space. Having to explore a fraction less than $10^{-29}$ of sequence space to find a specific new structure cannot possibly be bad for an evolutionary search.

This argument, however, has several problems. First, its flip side is that a $0.2n$-neighborhood is still astronomically large. Consider a hypothetical bacterial population of size $N=10^9$ individuals with a genic mutation rate of $10^{-6}$ per generation. Such a population would generate $10^3$ variants of the above RNA molecule per generation, and thus would need $1.43\times10^{28}$ generations to find most variants. For a bacterium like *Escherichia coli*, with approximately 300 generations per year in the wild (Ochman et al. 1999), this translates into $3.92\times10^{25}$ years, much longer than the age of the earth. Not even in vitro approaches, which can generate large populations of random molecules with a specified sequence can cover this small fraction of sequence space, as they are limited to some $10^{15}$ molecules (Wilson & Szostak 1999). Many biologically important RNA molecules, of course, are much longer than 100 nucleotides, which further aggravates the problem. A second problem is that shape space covering may be a peculiarity of RNA. Limited evidence suggests that it may not hold for proteins (Bornberg-Bauer 1997; Bornberg-Bauer 2002; Li et al. 1996; Nelson & Onuchic 1998; Todd et al. 1999). It may not hold either for other biological systems that can be represented in a discrete configuration space like sequence space. A case in point are generalized models of transcriptional regulation networks whose phenotypes do not show properties analogous to shape space covering (Ciliberti et al. 2006). A final and most serious problem with shape space covering is that it assumes that an evolutionary search can stray arbitrarily far from a given structure, and can thus freely explore any neighborhood of radius $0.2n$. This will not be the case if the starting structure needs to be preserved during this search.

All these reasons suggest that shape space covering may be less important for evolutionary innovation than one might think. In doing so, they also emphasize the importance of neutral networks for innovation.

**Acknowledgments**

This work was supported by EEC's FP6 Programme under contract IST-001935 (EVERGROW). The LPTMS is an Unite de Recherche de l'Universite Paris XI associee au CNRS. AW thanks the Swiss National Foundation for support

**Methods**

For our analyses, we used the Vienna RNA package (http://www.tbi.univie.ac.at/~ivo/RNA/ Hofacker et al. 1994), including the routines `fold`, which determines the minimum free energy (mfe) structure of a sequence; `inverse_fold`, which creates sequences folding into a given minimum free energy structure, using a guided random walk through sequence space that begins with a randomly chosen sequence (see also below); `pf_fold`, which computes a sequence's partition function (sum over all possible pairings, each weighted by the corresponding Boltzmann factor); `hamming`, which can calculate Hamming distances (number of mismatched bases) between two genotypes, as well as between two structures in their dot-parenthesis representation; and `bp_distance`, which determines an alternative measure of distance between two secondary structures, the bond distance. This distance is defined as the minimum number of base pairs that have to be opened or closed to transform one structure into the other.

*Generation of random sequences with a given mfe structure*

Uniform sampling of sequences that fold to a target structure cannot be efficiently done by generating sequences at random. Instead, we first generated sequences at random that are *compatible* with a target structure (forcing the bases that are paired in the target to be bases that *can* pair, i.e., A-U, C-G, and G-U), and determined for each such sequence whether it folded into the target structure. This approach is practical for sequences up to 50 base pairs. For longer sequences, one is forced to use heuristic approaches, such as the `inverse_fold` routine implemented in the Vienna RNA package (http://www.tbi.univie.ac.at/~ivo/RNA/ Hofacker et al. 1994). To justify using this routine, we studied its statistical bias by comparing its results to results obtained by random sampling of compatible sequences. Although we found that `inverse_fold` does not sample the desired space of sequences uniformly (sequences that are on the "boundary" of this space are sampled more frequently), this sampling bias is modest and would not qualitatively change our results. We hence used `inverse_fold` for our analyses. The routine occasionally fails to find a target structure from a given random sequence (e.g., 20% of the time for sequences of length 54), in which case we simply recommence with a new random sequence.

*Statistical exploration of k-neighborhoods*

Starting from a sequence $G_0$, we first identify all its neutral neighbors and take a step randomly to one of them, $G'$; we then determine all neutral neighbors of $G'$, take a step to one of them, and so on, thereby generating a random walk on the neutral network. This random walk ends at a pre-specified $k$ that is sufficiently large to collect the statistics that we are interested in. This procedure is repeated multiple times, i.e., the walk is started repeatedly at the same $G_0$, to obtain statistically meaningful results of its exploratory behavior as a function of $k$.

To study neutral "neighbors" at large genotype distances $k$ from a given sequence $G_0$, this procedure is computationally costly. In this case, we repeatedly generate pairs of genotypes folding into the structure at random, using `inverse_fold`, and analyze properties of these pairs as a function of their mutual distance $k$.

*Generic (random) structures*

We wished to assess whether observations made for our two biologically important structures also hold for generic structures. We define such structures as structures attained by randomly chosen genotypes. (All sequences we examined attain some mfe.) Specifically, for Figures 1 and 2, we produced 6000 random sequences $G_0$ and determined the distance between their 1-mutant neighbors and $P$. For all subsequent figures, we produced 100 such random structures, analyzed each of them in the way we analyzed the biological structures, and report results as averages over these 100 structures. We did this for structures of the same lengths (54 and 76) as our biological structures. (This means that for each random structure we generated 100 random $G_0$s, while for each biological structure we generate $10^4$ $G_0$s.) In both cases, `inverse_fold` is called enough times to produce $10^4$ successful hits to the target structures.

*Mutational robustness and thermodynamic stability*

To calculate the mutational robustness of a given sequence (genotype) *G*, we generate all of its 3*n* single base mutations (1-neighbours) and determine their mfe structure. The mutational robustness of *G* is then the number or the fraction of these neighbors that folds into the same mfe structure as *G*. As an indicator of the thermodynamic stability of the structure adopted by *G*, we determine the structural ensemble's free energy *F* using `pf_fold`, as well as the free energy $F_0$ of the mfe structure using `fold`. The thermodynamic stability of the mfe structure is defined as the probability $\exp[-(F_0-F)/kT]$ to find the sequence folded in its mfe structure when in thermodynamic equilibrium $\exp[-(F_0-F)/kT]$. In agreement with (Ancel & Fontana 2000), we observed that mutational robustness and thermodynamic stability are highly correlated. We therefore focused in our analysis on mutational robustness.

**Figure Captions**

**Figure 1:** Distribution of structure (Hamming) distances for structures in the 1-mutant neighborhood of sequences $G_0$ sampled at random (using inverse folding, Hofacker et al. 1994) from the neutral network of a sequence with phenotype $P$. **a)** 54nt hammerhead structure ((((((((.(((((...)))))........(((((......)))))...)))))))) involved in the self-cleavage of peach latent mosaic viroid (PMLVd Ambros et al. 1998); the most frequent variant (frequency 0.037) is .(((((((.(((((...)))))........(((((......)))))...))))))). **b)** 76nt tRNA$^{Phe}$ structure (((((((.(((((...)))))........(((((......)))))...))))))); the most frequent variant (frequency 0.0117) is (((((...((((........)))).(((((.....)))))).....(((((........))))).)))))).... Each distribution is obtained from 6000 $G_0$s (see Methods).

**Figure 2:** Distribution of structure (bond) distances for structures in the 1-mutant neighborhood of sequences $G_0$ sampled at random (using inverse folding Hofacker et al. 1994) from the neutral network of a sequence with phenotype $P$. **a)** 54nt hammerhead structure ((((((((.(((((...)))))........(((((......)))))...)))))))) involved in the self-cleavage of peach latent mosaic viroid (PMLVd Ambros et al. 1998). **b)** 76nt tRNA$^{Phe}$ structure ((((((((.(((((...)))))........(((((......)))))...)))))))). Each distribution is obtained from 6000 $G_0$s (see Methods).

**Figure 3**: $N_k$ is the set of structures in the 1-neighborhood of a sequence $G_k$ that do not also occur in the 1-neighborhood of $G_0$, where $G_k$ differs in $k$ nucleotides from the reference sequence $G_0$ and adopts the same structure $P$. See text for details.

**Figure 4**: Rapid accumulation of new variation with distance on a neutral network. The horizontal axes show the Hamming distance $k$ of a sequence $G_k$ to a sequence $G_0$, where both $G_0$ and $G_k$ adopt the same structure $P$. The vertical axes show the fraction of the structures in the 1-neighborhood of $G_k$ that lie in the set $N_k = I_k \setminus (I_0 \cap I_k)$. **a)** 54nt hammerhead structure involved in the self-cleavage of peach latent mosaic viroid (PMLVd Ambros et al. 1998); **b)** Average over 100 randomly generated structures with 54 bases. **c)** 76-mer tRNA cloverleaf structure. **d)** Average over 100 randomly generated

structures with 76 bases. All data was generated using inverse folding (Hofacker et al. 1994). Data as a function of $k$ were generated via the random walk approach described in Methods. Dots and bars represent mean ± standard errors.

**Figure 5:** Uniform random sampling of the neutral network leads to results similar to those when using inverse folding. The horizontal axes show the Hamming distance $k$ of a sequence $G_k$ to a sequence $G_0$, where both $G_0$ and $G_k$ adopt the same structure. The vertical axes show the fraction of structures in the 1-neighborhood of $G_k$ that lie in the set $N_k = I_k \setminus (I_0 \cap I_k)$. In contrast to Figures 4a and 4c, sequences $G_0$ folding into a given structure were not determined by inverse folding, but by uniform random sampling of the neutral network (see Methods). **a)** 54nt hammerhead structure involved in the self-cleavage of peach latent mosaic viroid (PMLVd Ambros et al. 1998). **b)** tRNA$^{Phe}$ cloverleaf structure. Data shown are averages over $10^4$ sequences $G_0$. Dots and bars represent mean ± standard errors.

**Figure 6**: The horizontal axes show the Hamming distance $k$ of a sequence $G_k$ to a sequence $G_0$, where both $G_0$ and $G_k$ adopt the same structure $P$. The vertical axes show the average pairwise *Hamming* distance $D_{I0, Ik}$ of all structures in the sets $I_0$ and $I_k$ (Figure 3), as well as the average distance $D_{P,Nk}$ of structures in $N_k$ to the structure $P$ itself. **a)** 54nt hammerhead structure involved in the self-cleavage of peach latent mosaic viroid (PMLVd Ambros et al. 1998). **b)** Average over 100 randomly generated structures with 54 bases. **c)** 76-mer tRNA cloverleaf structure. **d)** Average over 100 randomly generated structures with 76 bases. (See Methods for the construction of the $G_0$ and $G_k$.)

**Figure 7**: The horizontal axes show the Hamming distance $k$ of a sequence $G_k$ to a sequence $G_0$, where both $G_0$ and $G_k$ adopt the same structure $P$. The vertical axes show the average pairwise *bond* distance (see Methods) $D_{I0, Ik}$ of all structures in the sets $I_0$ and $I_k$ (Figure 3), as well as the average distance $D_{P,Nk}$ of structures in $N_k$ to the structure $P$ itself. **a)** 54nt hammerhead structure involved in the self-cleavage of peach latent mosaic viroid (PMLVd Ambros et al. 1998). **b)** Average over 100 randomly generated structures

with 54 bases. **c)** 76-mer tRNA cloverleaf structure. **d)** Average over 100 randomly generated structures with 76 bases. (See Methods for the construction of the $G_0$ and $G_k$ .)

**Figure 8:** The number of different new structures around a sequence declines with the sequences' mutational robustness. The horizontal axes show mutational robustness, the number of a sequence's $G_0$ neutral neighbors. The vertical axes show the mean number of structures in the 1-neighborhood of $G_0$ that are different from the structure $P$ of $G_0$. **a)** 54nt hammerhead structure involved in the self-cleavage of peach latent mosaic viroid (PMLVd Ambros et al. 1998). **b)** Average over 100 randomly generated structures with 54 bases. **c)** tRNA$^{Phe}$ cloverleaf structure. **d)** Average over 100 randomly generated structures with 76 bases. (See Methods for the construction of sequences $G_0$ .)

**Figure 9.** Dependency of $D_{I0, Ik}$ and $D_{G0,Nk}$ on the genotype distance $k$, and on the mutational robustness of $G_0$ for the tRNA$^{Phe}$ cloverleaf structure. The lower-right panel shows the distribution of mutational robustness for the reference sequences used in this analysis. The other panels contain data analogous to Figure 6, but where reference sequences are categorized according to their mutational robustness $R$ (as indicated in each panel). Robustness $R$ is defined as the number of a sequence's immediate neighbors that adopt the same structure. (See Methods for the construction of sequences $G_0$ and $G_k$ .)

.

**Literature Cited**

Ambros, S., Hernandez, C., Desvignes, J. & Flores, R. 1998 Genomic structure of three phenotypically different isolates of peach latent mosaic viroid: Implications of the existence of constraints limiting the heterogeneity of viroid quasispecies. *Journal of Virology* 72, 7397-7406.

Ancel, L. W. & Fontana, W. 2000 Plasticity, evolvability, and modularity in RNA. *Journal of Experimental Zoology/Molecular Development and Evolution* 288, 242-283.

Birikh, K., Heaton, P. & Eckstein, F. 1997 The structure, function, and application of the hammerhead ribozyme. *European Journal of Biochemistry* 245, 1-16.

Bornberg-Bauer, E. 1997 How are model protein structures distributed in sequence space? *Biophysical Journal* 73, 2393-2403.

Bornberg-Bauer, E. 2002 Randomness, structural uniqueness, modularity and neutral evolution in sequence space of model proteins. *Zeitschrift fur Physikalische Chemie - International Journal of Research in Physical Chemistry & Chemical Physics* 216, 139-154.

Ciliberti, S., Martin, OC. & Wagner, A. 2006 Innovation and robustness in complex regulatory gene networks (submitted).

Fontana, W. 2002 Modelling 'evo-devo' with RNA. *Bioessays* 24, 1164-1177.

Fontana, W. & Schuster, P. 1998a Continuity in evolution: On the nature of transitions. *Science* 280, 1451-1455.

Fontana, W. & Schuster, P. 1998b Shaping space: the possible and the attainable in RNA genotype-phenotype mapping. *Journal of Theoretical Biology* 194, 491-515.

Higgs, P. 1993 RNA secondary structure: a comparison of real and random sequences. *J. Phys. I France* 3, 43-59.

Hofacker, I., Fontana, W., Stadler, P., Bonhoeffer, L., Tacker, M. & Schuster, P. 1994 Fast folding and comparison of RNA secondary structures. *Monatshefte fuer Chemie* 125, 167-188.

Huynen, M. 1996 Exploring phenotype space through neutral evolution. *Journal of Molecular Evolution* 43, 165-169.

Huynen, M., Stadler, P. & Fontana, W. 1996 Smoothness within ruggedness: The role of neutrality in adaptation. *Proceedings of the National Academy of Sciences of the U.S.A.* 93, 397-401.

Li, H., Helling, R., Tang, C. & Wingreen, N. 1996 Emergence of preferred structures in a simple model of protein folding. *Science* 273, 666-669.

Nelson, E. & Onuchic, J. 1998 Proposed mechanism for stability of proteins to evolutionary mutations. *Proceedings of the National Academy of Sciences of the U.S.A.* 95, 10682-10686.

Ochman, H., Elwyn, S. & Moran, N. A. 1999 Calibrating bacterial evolution. *Proceedings of the National Academy of Sciences of the U.S.A.* 96, 12638-12643.

Reidys, C., Stadler, P. & Schuster, P. 1997 Generic properties of combinatory maps: Neutral networks of RNA secondary structures. *Bulletin of Mathematical Biology* 59, 339-397.

Schuster, P., Fontana, W., Stadler, P. & Hofacker, I. 1994 From sequences to shapes and back - a case-study in RNA secondary structures. *Proceedings of the Royal Society of London Series B* 255, 279-284.

Tacker, M., Stadler, P., BornbergBauer, E., Hofacker, I. & Schuster, P. 1996 Algorithm independent properties of RNA secondary structure predictions. *European Biophysics Journal with Biophysics Letters* 25, 115-130.

Todd, A., Orengo, C. & Thornton, J. 1999 Evolution of protein function, from a structural perspective. *Current Opinion in Chemical Biology* 3, 548-556.

van Nimwegen, E., Crutchfield, J. & Huynen, M. 1999 Neutral evolution of mutational robustness. *Proceedings of the National Academy of Sciences of the U.S.A.* 96, 9716-9720.

Wilson, D. S. & Szostak, J. W. 1999 In vitro selection of functional nucleic acids. *Annual Review of Biochemistry* 68, 611-647.

Zuker, M. 2000 Calculating nucleic acid secondary structure. *Current Opinion in Structural Biology* 10, 303-310.

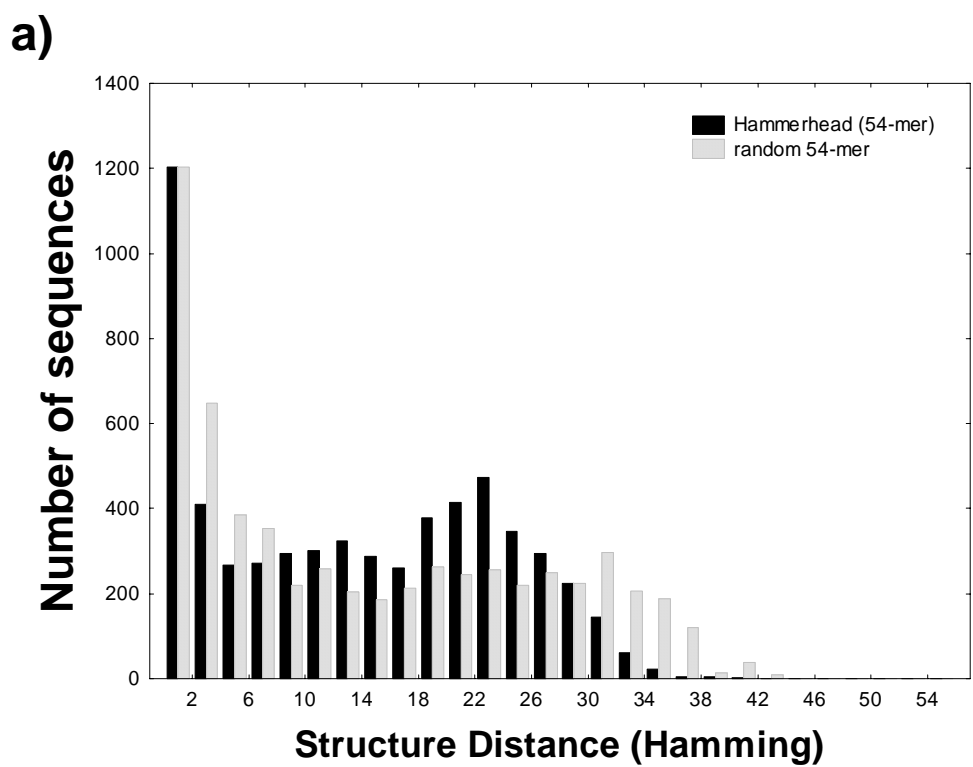
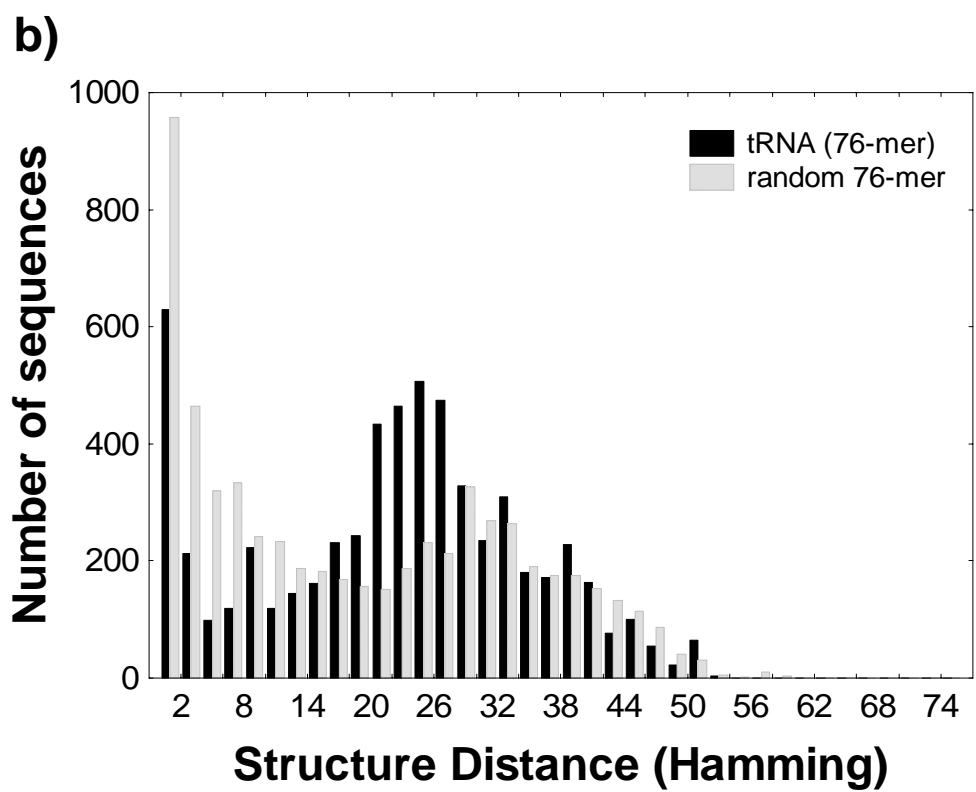

**Figure 1**

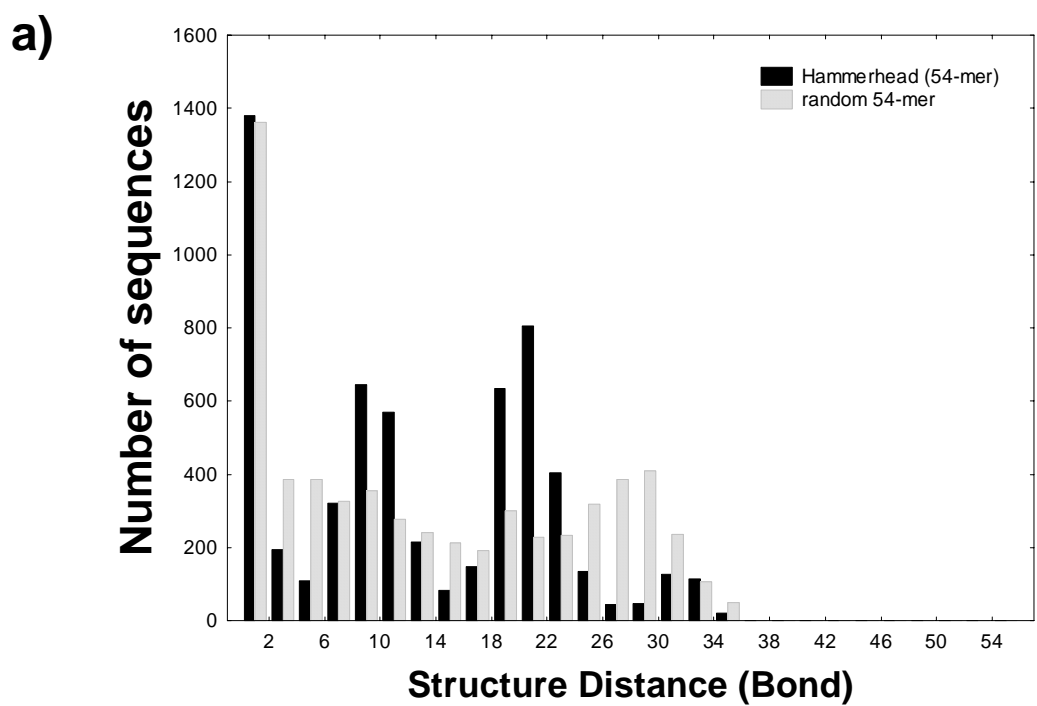
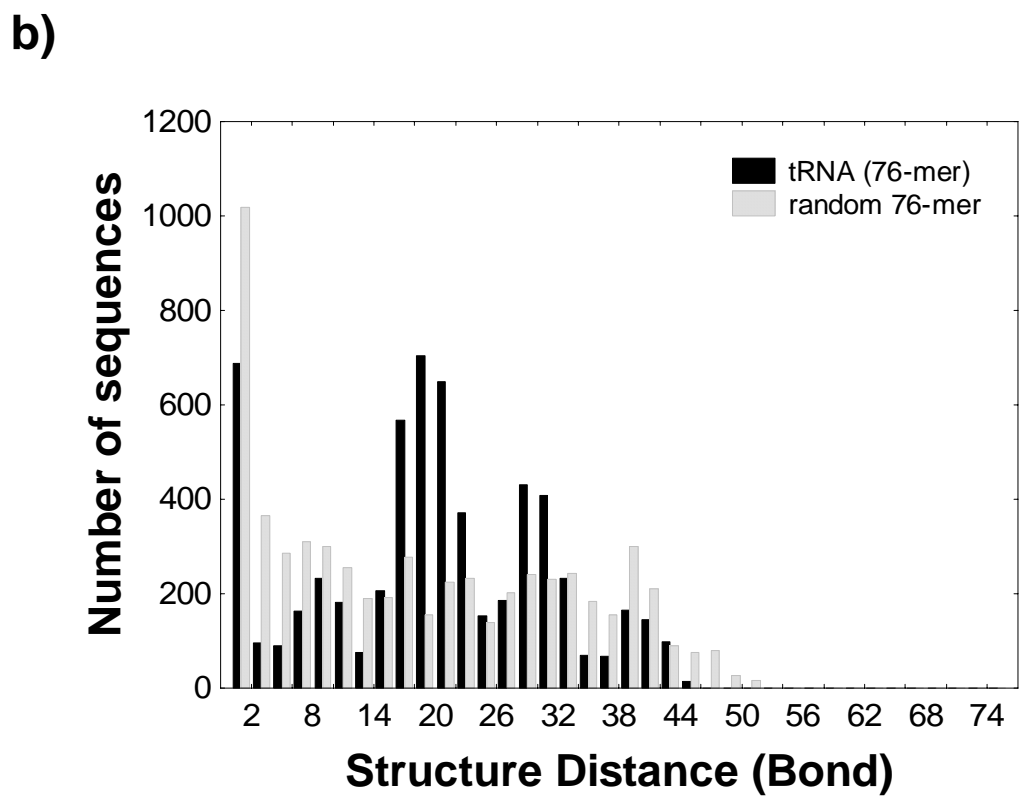

**Figure 2**

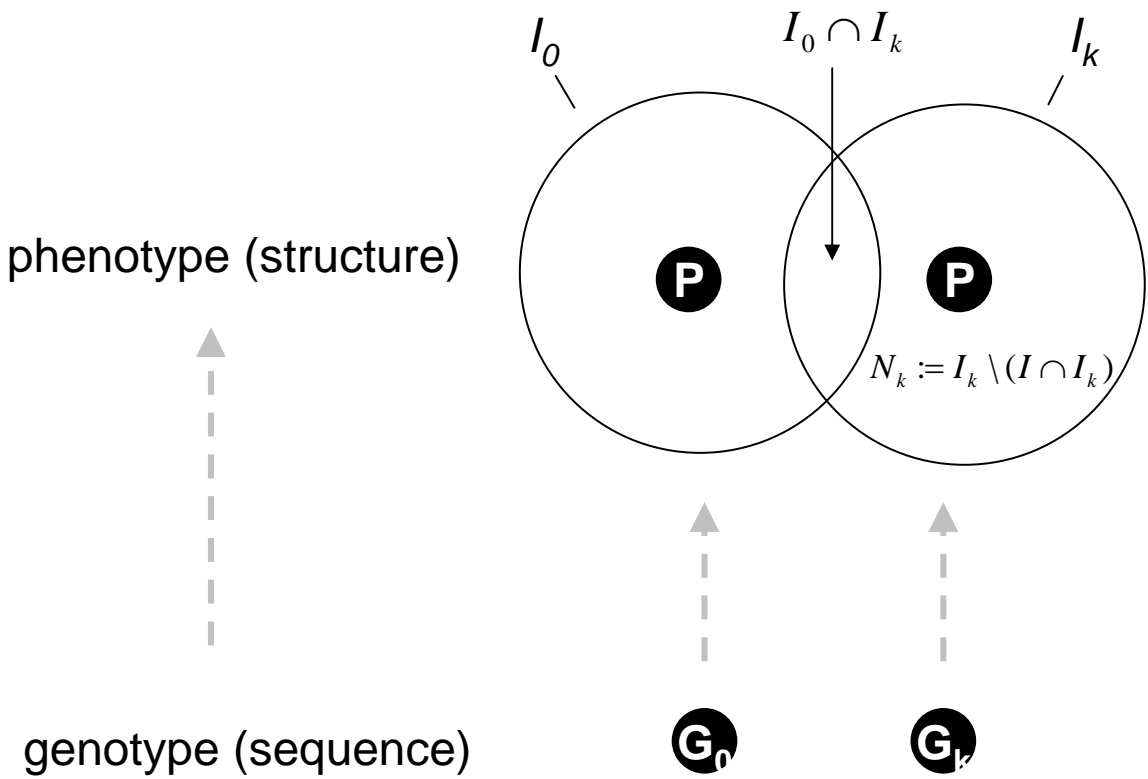

**Figure 3**

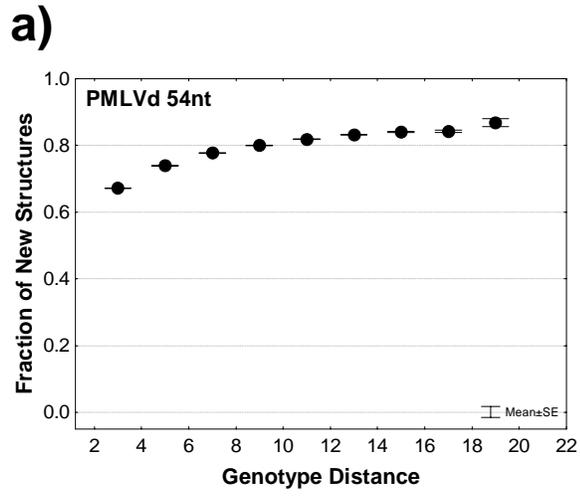
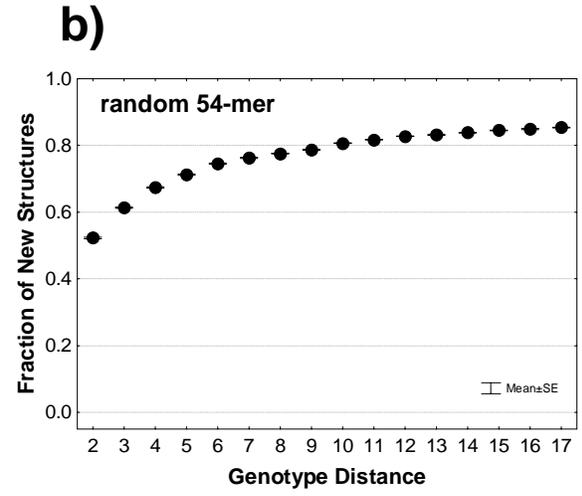
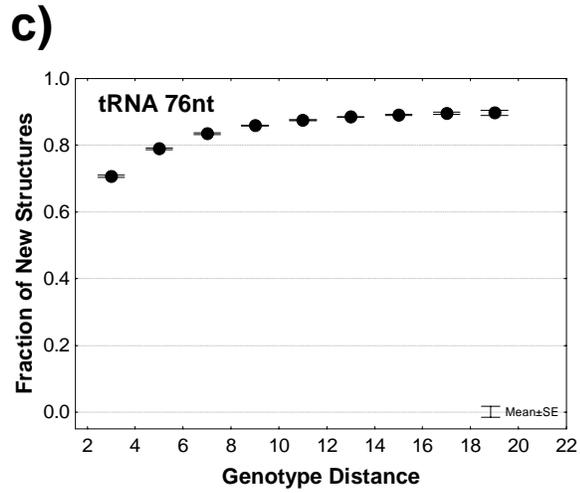
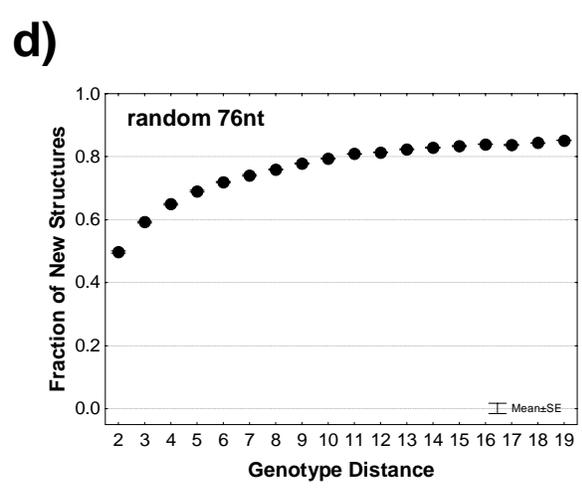

**Figure 4**

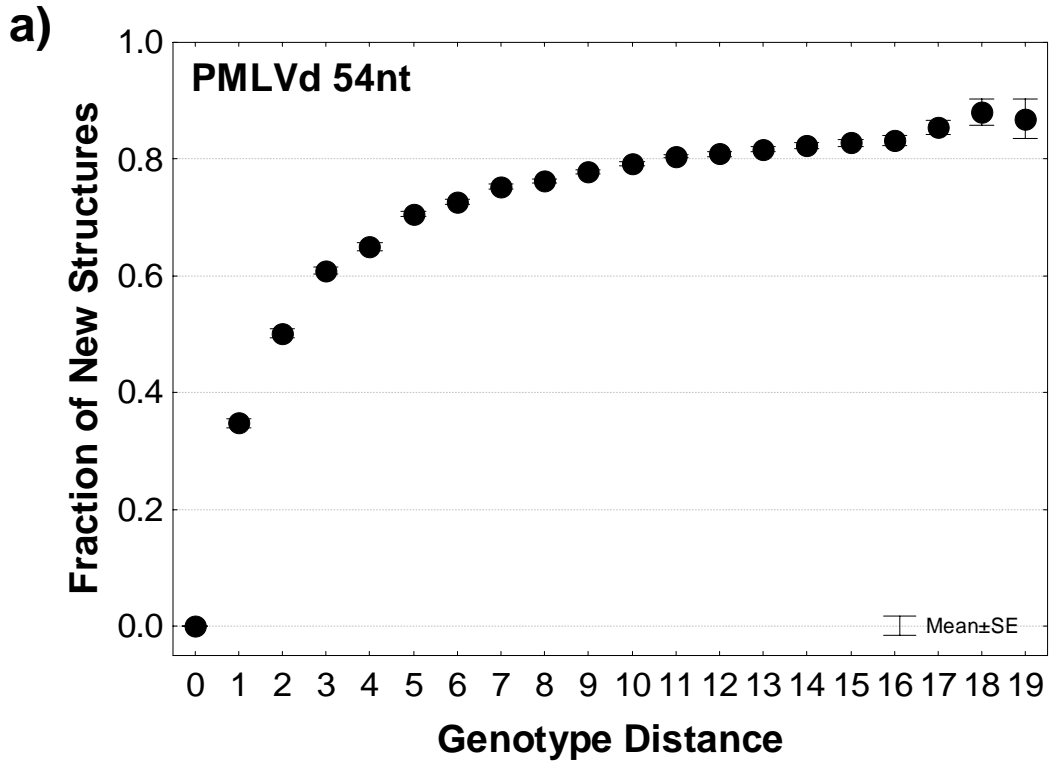
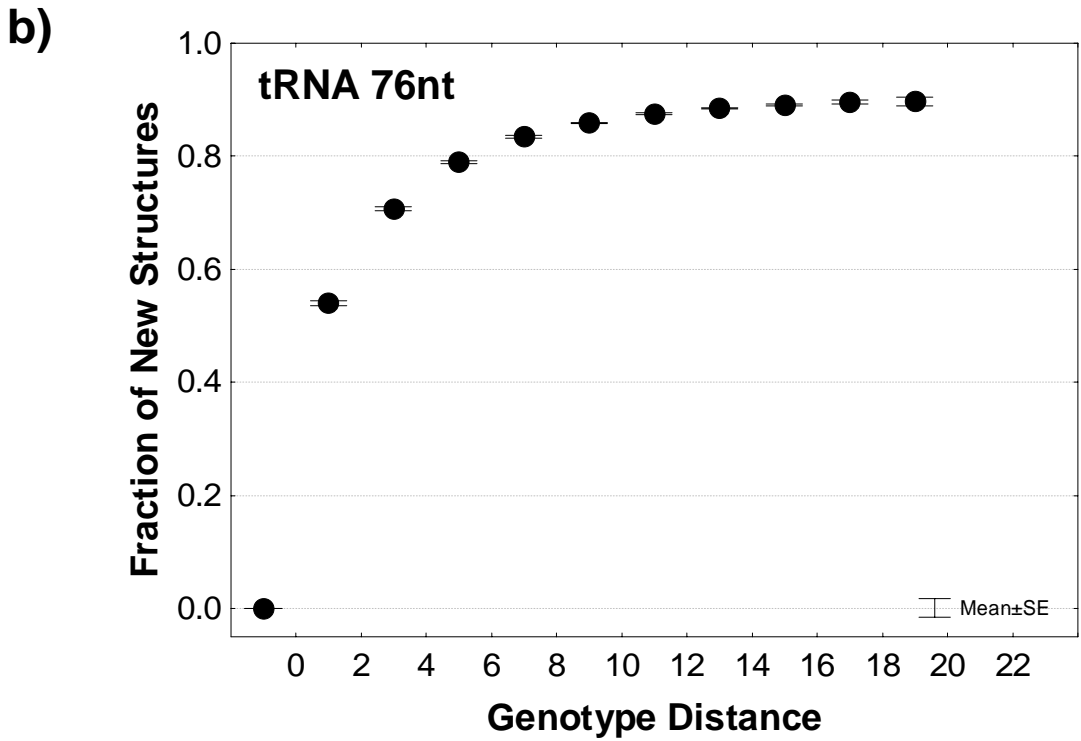

**Figure 5**

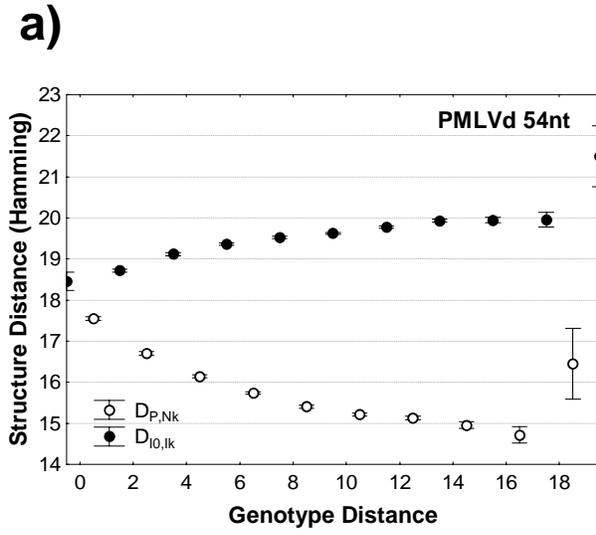
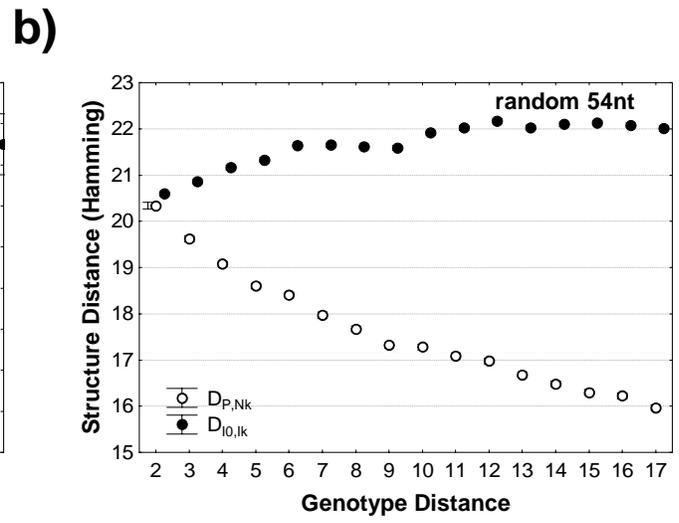
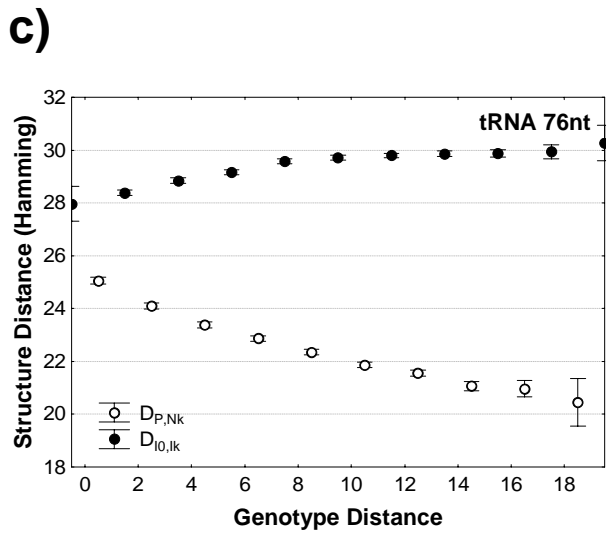
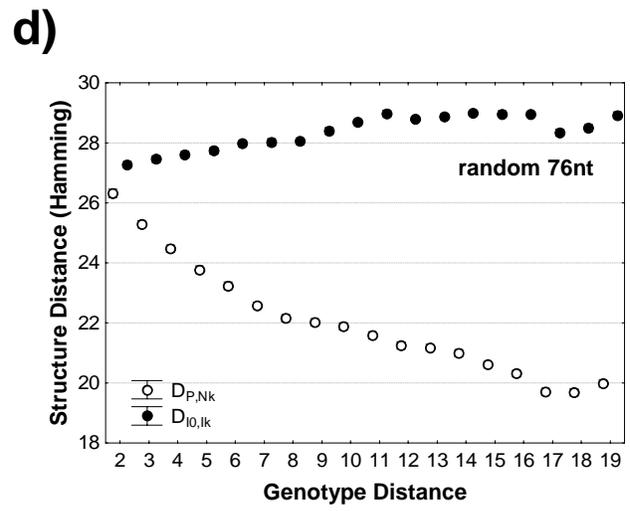

**Figure 6**

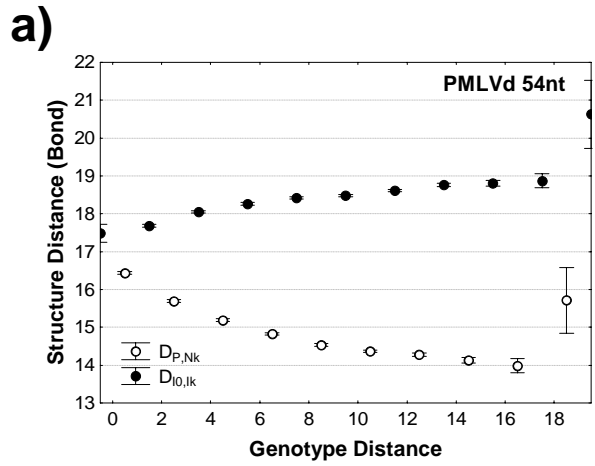
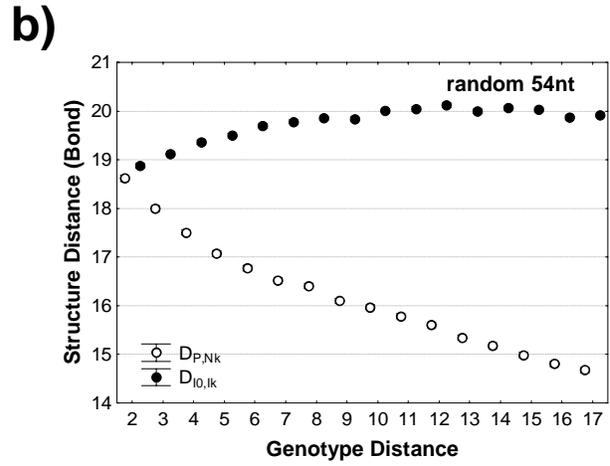
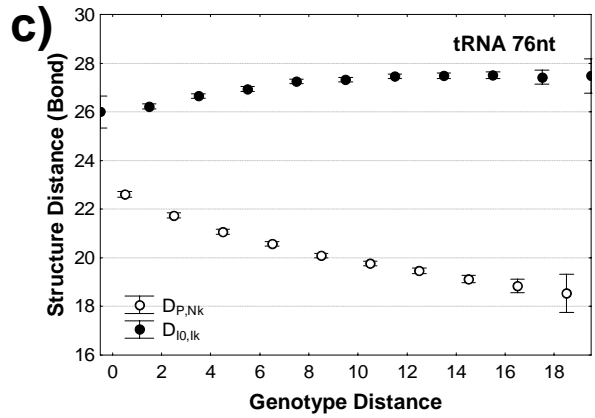
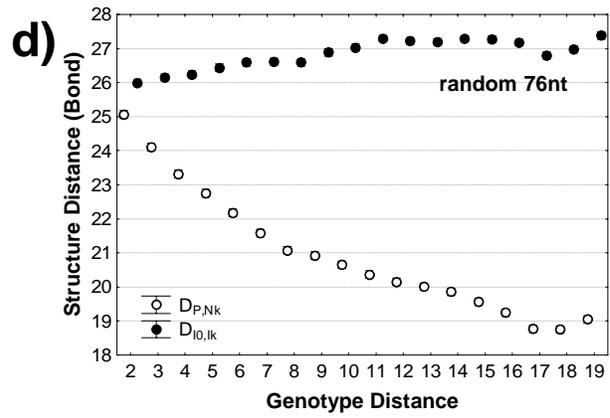

**Figure 7**

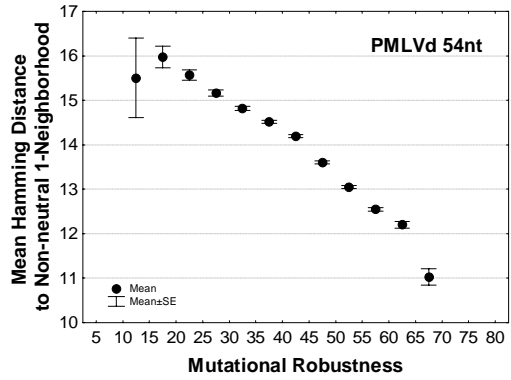
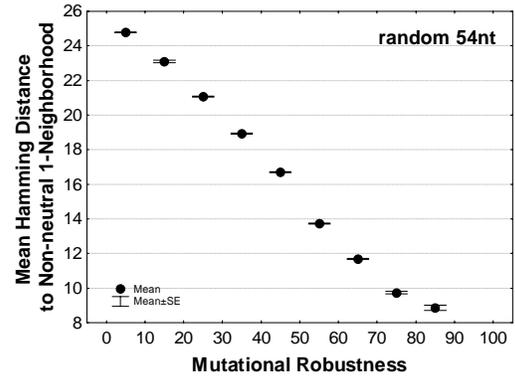
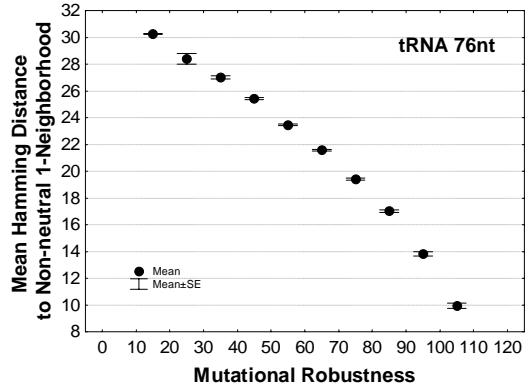
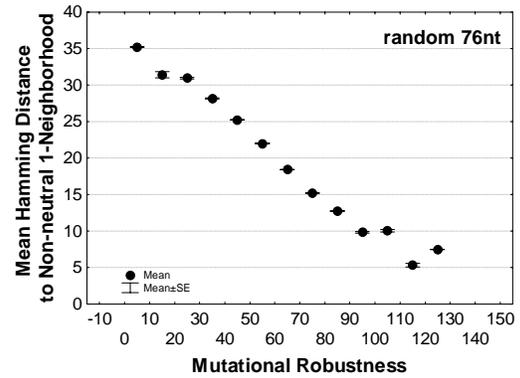

**Figure 8**

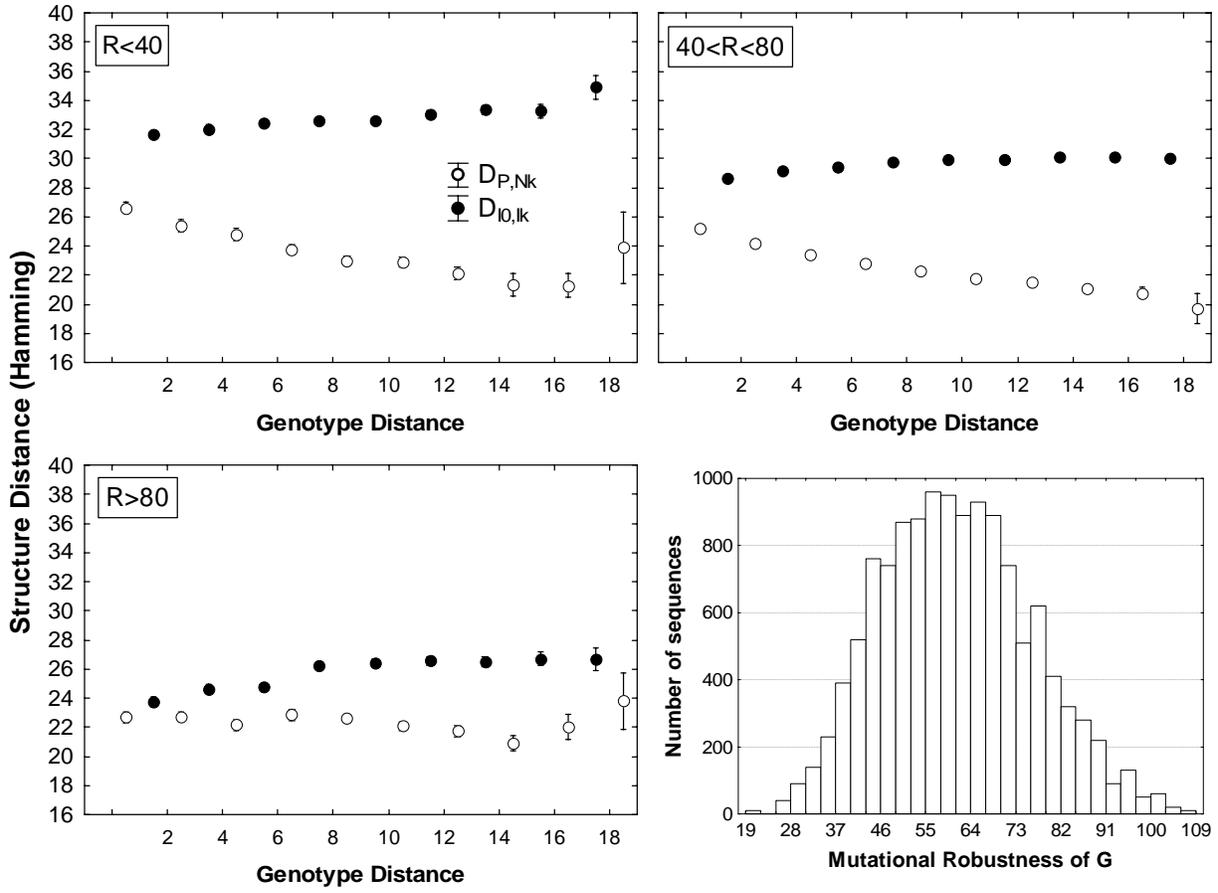

**Figure 9**